# Automated Vulnerability Detection Using Deep Learning Technique

**Guan-Yan Yang[1,*], Yi-Heng Ko[1], Farn Wang[1], Kuo-Hui Yeh[2], Haw-Shiang Chang[1] and Hsueh-Yi Chen[1]**

[1]Department of Electrical Engineering, National Taiwan University, Taipei, 106319, Taiwan R.O.C.

[2]Institute of Artificial Intelligence Innovation, National Yang Ming Chiao Tung University, Hsinchu City 300093, Taiwan R.O.C.

[*]Corresponding Author: Guan-Yan Yang. Email: f11921091@ntu.edu.tw

## ABSTRACT

## 1 Introduction

Ensuring the absence of exploitable vulnerabilities within applications has always been a critical aspect of software development [1-3]. Traditional code security testing methods often rely on manual inspection or rule-based approaches, which can be time-consuming and prone to human errors. With the recent advancements in natural language processing, deep learning has emerged as a viable approach for code security testing. In this work, we investigated the application of deep learning techniques to code security testing to enhance the efficiency and effectiveness of security analysis in the software development process. In 2022, Wartschinski et al. [4] utilized a Word2Vec model for Python source code embedding, followed by a Long Short-Term Memory (LSTM) [5] model to identify vulnerable code patterns. Building upon their research, we evaluated the performance of two embedding methods in generating code vector representations to improve training efficiency. By leveraging the CodeBERT [6] model instead of Word2Vec, we achieved enhancements in detecting SQL injection vulnerabilities [7]. Additionally, we applied our proposed models to multiple projects collected from GitHub, compared the scan results with existing static testing tools, and evaluated their performance. The results indicate that our approach outperforms commercially available static application security testing (SAST) tools [8-9], showcasing the potential of deep learning in advancing code security testing.

## 2 Methodology

In this section, we present our approach. Figure.1 is an overview of our method. We collected the vulnerability dataset from Github, and adopted CodeBERT [6] as the embedding method. After transferring the source code into a vector representation, we trained an LSTM model to extract the vulnerable pattern from the code.

## 2.1 Data Collection and Labeling

We chose SQL injection vulnerability [7] as our detecting objective since it is reported as one of the most common vulnerabilities by OWASP (https://owasp.org/Top10) and CVE databases (https://cve.mitre.org/). On top of that, SQL injection has syntactic features that can be learned by a deep learning model. To collect the SQL injection dataset from GitHub, we employ the PyDriller [10]. We apply keywords to filter the commits we want, such as commit messages containing "SQL injection fixed" or "SQL injection prevented," etc. We consider the changed part as it potentially would cause SQL injection. Therefore, we mined the commit and marked the changed part as vulnerable.

## 2.2 Embedding Layers

CodeBERT is a bimodal pre-trained model for NL-PL(Natural Language-Programming Language) tasks, which uses a multi-layer bidirectional Transformer as the model architecture, training at a large dataset for six different programming languages and natural language. Different from [4], we use CodeBERT for Python source code embedding and feed all output tokens to an LSTM model.

## 2.3 LSTM Model Training

After pre-processing the data, we can start training our LSTM model. To implement the LSTM model, we used the Keras library. There are three layers in the LSTM architecture: LSTM layer, dropout layer, and dense layer. To yield a better performance, the hyper-parameters setting of the model must be carefully chosen. Table.1 shows the hyper-parameters of the LSTM model. Due to space constraints, we will not discuss the details of parameter settings.

## 2.4 Result

Table.2 illustrates the model performance on detecting SQL injection vulnerabilities within source code. One thing that has to be mentioned is that the accuracy here is reported only for completeness reasons. Due to the data imbalance, most of the code will be clean, and vulnerable parts are relatively rare, meaning there are many more negative parts than positive parts. We can see that our model yields a satisfying result on vulnerability detection by reaching 86.2% precision, 80.0% recall, and 83.1% f1-score.

## 3 Evaluation

To analyze performance, we used the model architectures in [4] and trained them on our dataset. The results are presented in Table.2. Our observation indicates that our model outperforms the research of [4], attributed to CodeBERT's utilization of a self-attention mechanism, facilitating better retrieval of contextual information compared to Word2Vec. Furthermore, CodeBERT is trained with natural language concurrently. In contemporary software development, there is a notable emphasis on writing human-readable source code, involving the use of meaningful names for functions and variables, as well as documenting code in natural language [11]. Therefore, CodeBERT has better results in this scenario.

To further evaluate our model in real-world scenarios, we curated an additional dataset focusing on SQL injection vulnerabilities and tested it with SAST tools, namely Bandit [8] and Checkmarx [9]. Bandit is an open-source tool for Python source code vulnerability detection, while Checkmarx is a commercial tool offering a comprehensive suite of application security testing, including SAST. Table.3 presents the testing results on 97 Python files potentially containing SQL injection vulnerabilities. Our model exhibited superior performance, attributable to SAST tools often relying on predefined features crafted by humans. However, given the diverse forms of vulnerabilities, human-defined features may not cover all vulnerable modes. In contrast, our method learns from large amounts of data, allowing it to capture semantics and predict vulnerabilities more effectively.

## 4 Conclusion

In this work, we proposed a method for Python SQL injection vulnerability detection. By reproducing previous work, we found that CodeBERT is more suitable than Word2Vec for source code embedding tasks. After that, we compare our model with two SAST tools in real-world cases, and the result shows that the deep learning model has the potential to win against traditional static analysis tools on vulnerability detection. In addition, our model architecture can further expand to other vulnerabilities and even other languages. Also, we are conducting more experiments to ensure that our method can detect more different types of vulnerabilities.

## KEYWORDS

Static Analysis; Deep Learning; Embedding; Vulnerability Detection;

**Figure 1**: Overview of approach

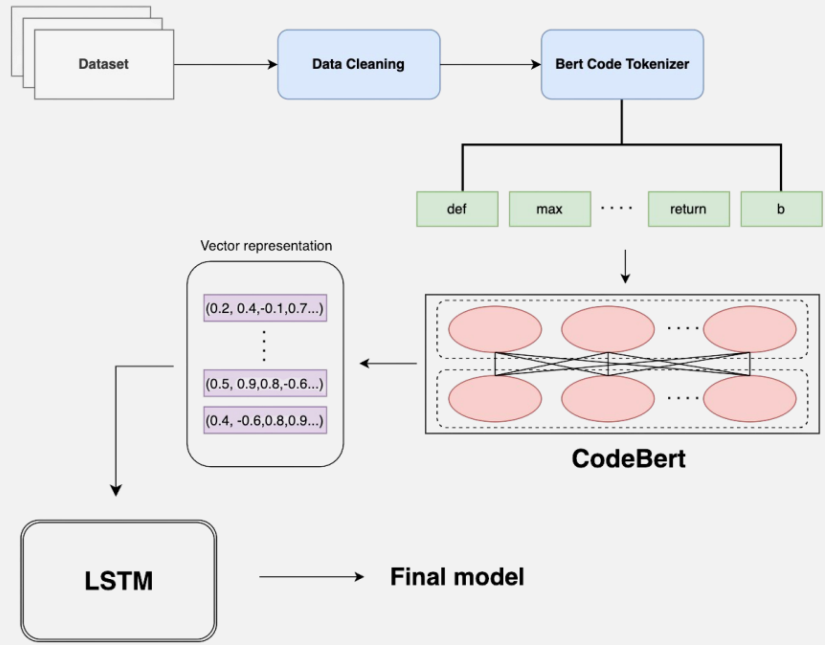

**Table 1**: Hyper-parameters of LSTM model

| Training parameters | Value |
| --- | --- |
| Epochs | 100 |
| Batch size | 128 |
| Neurons | 100 |
| Dropout | 20% |
| Learning rate | le-3 |
| Optimizer | Adam |

**Table 2**: Comparison of the results

| Name | Approach | Accuracy | Precision | Recall | F1 |
| --- | --- | --- | --- | --- | --- |
| Wartschinski et al. [3] | Word2Vec + LSTM | 92.5% | 82.2% | 78.0% | 80.1% |
| Our model | CodeBERT + LSTM | 92.5% | **86.2%** | **80.0%** | **83.1%** |

**Table 3**: Scanning results on real world project with SQL injection

| | Acc | Precision | Recall | F1 |
| --- | --- | --- | --- | --- |
| Our model | **93.1%** | **98.0%** | **94.2%** | **96.1%** |
| Bandit [4] | 78.9% | - | 76.9% | 87.0% |
| Checkmarx [5] | 88.3% | 94.1% | 92.3% | 93.2% |

**Funding Statement:** This work was partly supported by the National Science and Technology Council (NSTC) of Taiwan under grants MOST 110-2221-E-002-069-MY3, NSTC 112-2634-F-011-002-MBK, NSTC 113-2634-F-011-002-MBK, and NSTC 111-2221-E-A49-202-MY3. Additionally, it was supported by the 2024 CITI Visible Project: Questionnaire-based Technology for App Layout Evaluation, Academia Sinica, Taiwan, ROC. Moreover, this work received partial funding from the National Taiwan University under Grant 113L7256, within the framework of the Higher Education Sprout Project by the Ministry of Education, Taiwan. Guan-Yan Yang is supported by National Science and Technology Council Graduate Research Fellowship (NSTC-GRF), under Grant 113WFA0110119.

**Acknowlegements:** Thanks to Security Researcher Zhao Min Chen from CyCraft Technology, Taiwan, for his invaluable suggestions.

**Conflicts of Interest:** The authors declare that they have no conflicts of interest to report regarding the present study.